\documentclass[aps,pra,reprint]{revtex4-2}

\usepackage{mathtools,amssymb,amsmath,commath}
\usepackage{graphicx}
\usepackage{dcolumn}
\usepackage{bm}
\usepackage[colorlinks=true,bookmarks=true,allcolors=blue,citecolor=red,urlcolor=blue]{hyperref}

\DeclareMathOperator{\sech}{sech}

\usepackage{tikz}
\newcommand*\circled[1]{\tikz[baseline=(char.base)]{
		\node[shape=circle,draw,inner sep=0.2pt] (char) {#1};}}

\begin{document}
	
	\title{Bistable quartic soliton in saturable nonlinear media}
	\author{Tiyas Das}
	\author{Anuj Pratim Lara}
	\author{Samudra Roy }
	\email{samudra.roy@phy.iitkgp.ac.in}
	
	\affiliation{Department of Physics, Indian Institute of Technology Kharagpur, Kharagpur, India, 721302.
	}

	\begin{abstract}
		This report presents a theoretical demonstration of a novel bistable quartic soliton (BQS) in saturable nonlinear media, specifically within a realistic dispersion-engineered ridge waveguide made of Lithium Niobate (\( \text{LiNbO}_3 \)). The study employs the variational method to establish the amplitude-width relationship, indicating the coexistence of stable solitons with the same duration but differing amplitudes.  The impact of shock on the bistable soliton is examined through perturbative variational analysis, supported by numerical results. Additionally, we examine the interaction of BQS in different regimes and analyze the formation of the bound state. The robustness of the BQS under perturbations is further investigated via linear stability analysis.
	\end{abstract}
	
	\maketitle

	\section{Introduction}
	An optical soliton is a stable wave that retains its shape during long-distance propagation, achieved through the balance of group-velocity dispersion (GVD) and Kerr nonlinearity \cite{agrawal2000nonlinear}. 
	In addition to traditional solitons, a new class known as \textit{quartic soliton} (QS) was introduced in early 90s \cite{hook1993ultrashort,karlsson1994soliton,akhmediev1994radiationless,buryak1995stability}. 
	QS emerges in a specific dispersion regime where the coefficients of 3$^{rd}$-order dispersion ($\beta_3$) is zero, and both 2$^{nd}$ ($\beta_2$) and $4^{th}$ ($\beta_2$) order dispersion are negative \cite{piche1996bright}. This unique point on the dispersion curve is referred to as the \textit{quartic point} and essentially the inspiration to the author to coin the term \textit{quartic soliton} \cite{roy2013formation}. This new soliton type differs from conventional Kerr-solitons by having a secant hyperbolic square \cite{karlsson1994soliton} pulse shape and may exhibit radiationless oscillatory tails contingent on its propagation constant \cite{buryak1995stability}.

	After its first theoretical introduction in the 90s, research on QS has largely stagnated due to the adverse effects of Raman-induced frequency redshift. This redshift can cause the soliton frequency to shift away from the quartic point, where QS becomes non-existent. Notably, the Raman-induced frequency redshift ($\Delta \omega_R$) is significantly impacted by pulse duration ($t_0$), scaling as $\Delta \omega_R \propto t^{-6}_0$ for QS, compared to the conventional soliton's scaling of $\Delta \omega_R \propto t^{-4}_0$ for short pulses  in optical fibers \cite{wang2022raman}.
	Achieving the desired dispersion profile for QS has been a significant challenge; however, advancements in nanotechnology for waveguide fabrication have enabled the customization of dispersion characteristics.
	Exploiting such technique, recent theoretical and experimental studies have focused  \cite{blanco2016pure,kruglov2018solitary,tam2019stationary,taheri2019quartic,runge2020pure,Lo2018} specifically on the sub-branch of QS which is formally known as \textit{pure quartic soliton} (PQS) \cite{blanco2016pure, de2021pure}. PQSs are generated through the interaction of negative quartic dispersion and Kerr nonlinearity under conditions of zero GVD. Unlike Kerr solitons, which follow the energy-width relationship \(E \propto t^{-1}_0\), PQSs exhibit an advantageous energy scaling of \(E \propto t^{-3}_0\)  \cite{tam2019stationary}, indicating they can transport more energy for the low pulse duration \cite{Lo2018}. This characteristic positions PQSs as promising candidates for high-power laser applications \cite{runge2020pure}.

	Though the investigation of QS has been studied in various modulated dispersion environments \cite{runge2021infinite,tsoy2024generic,Runge2021}  its exploration is limited within the domain of Kerr nonlinearity. Perhaps the reason behind this is, QSs, PQSs, and conventional Kerr solitons are categorized as members of the same family \textit{generalized-dispersion Kerr solitons} \cite{Tam2020}.
	This limitation raises interest in exploring the existence of QS in broader types of nonlinear systems.  
	Motivated by this idea, in this report, we focus on investigating the formation of QS in an optical medium offering \textit{saturable nonlinearity} (SN) which, to the best of our knowledge, has never been explored before.

	The SN is typically associated with materials such as \textit{photo-refractive} (PR) \cite{bian1997photorefractive,christodoulides1995bright,lin2013ground} crystals and \textit{semiconductor doped glass} (SDG) \cite{coutaz1991saturation}. In this analysis, we focus on Lithium Niobate (LiNbO$_3$, LN), recognized for its significant PR properties, including a high  $\chi^{(3)}$ nonlinearity of $n_2=2.5\times 10^{-19}$ m$^2$/W and a broad optical transparency range (350 nm - 4500 nm) \cite{hamrouni2024picojoule}.
	The PR nonlinearity in LiNbO$_3$ arises from the light-induced space charge field $E_\text{sc}$, which modulates the refractive index through the electro-optic effect \cite{crosignani1998nonlinear}, contingent upon appropriate bias and orientation of the sample. Experimental evidence demonstrates that self-focusing SN can occur in LiNbO$_3$ under an external electric field aligned with the optic axis of the crystal \cite{christodoulides1995bright}. Notably, this PR effect is carrier-dependent, resulting in a delayed response compared to the Kerr effect when altering the refractive index. Effective modulation in PR materials is achieved through the application of high-irradiance femtosecond laser pulses, which facilitate the generation of SN \cite{gamaly2010modification,juodkazis2008laser,gamaly2008three}.

\begin{figure}[h!]
	\centering
	\includegraphics[width=\linewidth]{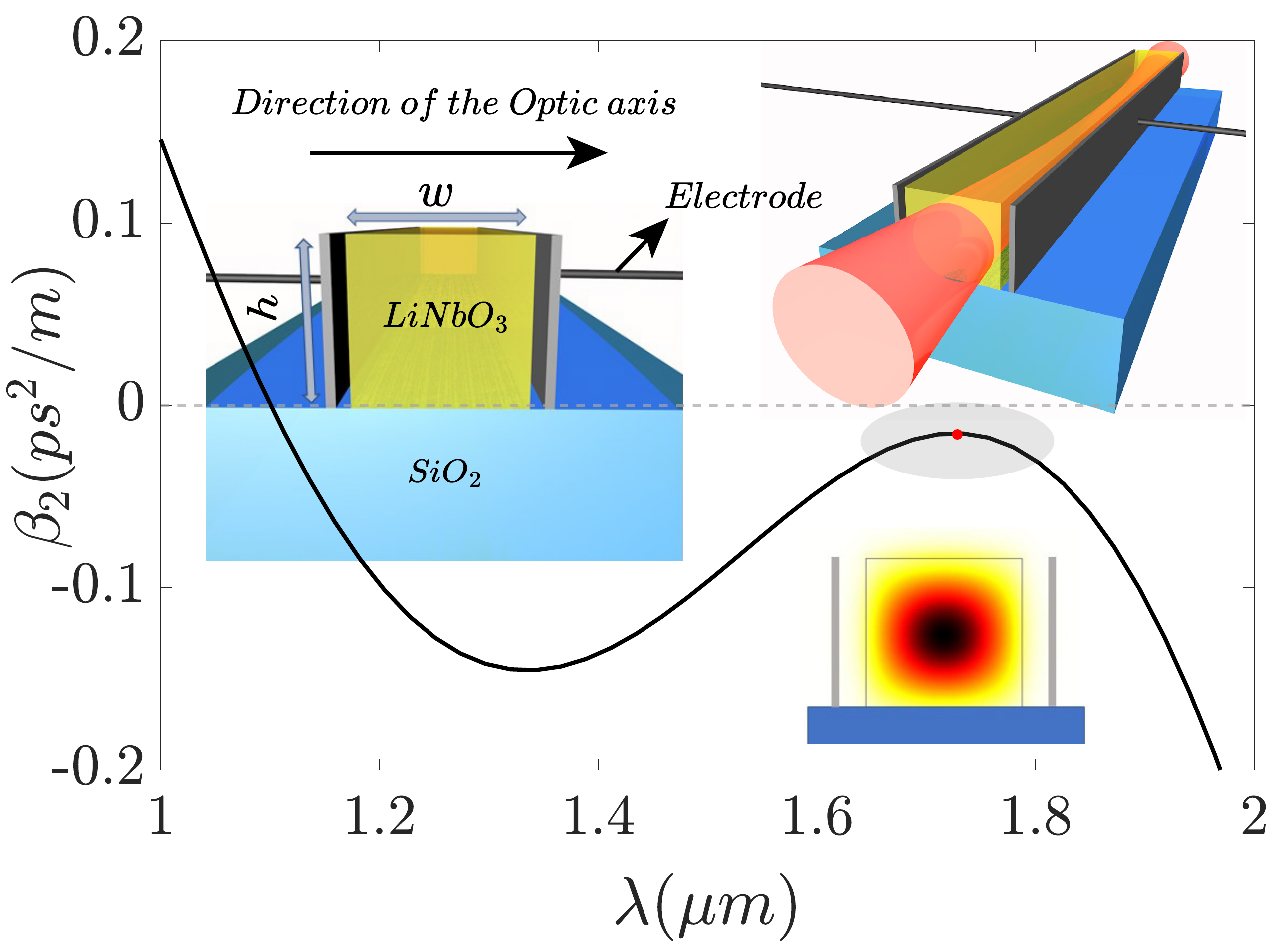}
	\caption{(GVD profile for a quasi-TE mode of the $LN-SiO_2$
		ridge-waveguide. The red dot on the GVD profile indicates the $\lambda_0$ at which 3OD vanishes ($\beta_3 = 0$) referred as the \textit{quartic point} (QP) and the shaded region around QP
		indicates the range where the approximation $\beta_3 \approx 0$ is valid. Inset below (left): The mode confinement at $ \lambda_0 = 1.73 \mu m$. Inset above (left): The front-view representation of the waveguide geometry where the parameters are
		$w = 2.2 \mu m$ and $h =2.1 \mu m$. Inset above (right): The schematic representation of the full set up. }
	\label{fig1}
\end{figure}

In this report, we investigate the existence of bistable quartic-solitons (BQSs) within a LiNbO$_3$ based waveguide that exhibits SN. Our theoretical analysis demonstrates that these robust BQSs can arise in suitable dispersion conditions. 
We examine the temporal dynamics of the BQSs under shock effects, which act as perturbations specially for SDGs. { Additionally, we investigate the interaction of BQS exhibiting oscillatory and non-oscillatory decaying tail. A force analysis ensures QS with oscillatory tail may form a bound state. Finally, we assess the stability of these BQSs against input amplitude noise and establish a stability map across a parametric space  includes incident wave amplitude and strength of the saturable nonlinearity. In actual experiments such parameters can be tuned through the input intensity and external bias field}.  Our findings indicate a pathway to realizing a new class of solitary waves in mediums with SN, which could be significant for applications in fiber lasers and optical communications. {In particular, these BQSs can be a potential candidate to build switching and logic-gate devices. Advantageous energy scaling further makes them a viable tool for bistable lasers. The tunability of the bistable curve through the saturable parameter provides additional flexibility in laser applications.}

\section{Theory and model}
We design a realistic ridge waveguide composed of LiNbO$_3$, featuring external bias along a transverse direction. The GVD profile for a quasi-transverse-electric (TE) mode of the extraordinary ray is illustrated in Fig.~\ref{fig1}. To excite SN, an external electric-field bias is applied along the optic axis \cite{christodoulides1995bright}, with uniform irradiation to avoid oversaturation \cite{bian1997photorefractive}. The waveguide cross-section, system set-up, and mode confinement are illustrated as an inset (upper and lower) in Fig.~\ref{fig1}.  The GVD-profile exhibits the \textit{quartic point} (red dot) encircled by a shaded elliptical region where the approximation  $\beta_3 \approx 0$ is valid and the GVD profile can be expressed by including only $\beta_2$ and $\beta_4$, terms respectively.   
In the proximity of the \textit{quartic point}, GVD is anomalous ($\beta_2 < 0$) and GVD curvature is negative ($\beta_4 < 0$). Exactly at the \textit{quartic point}, for operating wavelength $ \lambda_0 \approx 1.73 \mu$m  the dispersion coefficients are calculated as $\beta_2 = -0.014$ ps$^2$/m, $\beta_3 = 2.51\times 10^{-5}$ ps$^3$/m and $\beta_4 =- 1.30 \times 10^{-5} $ ps$^4$/m, respectively. 
Note that, for focusing nonlinearity, these are the primary conditions for the exploration of localized QS and we carefully design the waveguide structure to achieve the correct sign and curvature of the GVD at operating wavelength. 

The optical pulse propagation in the waveguides where the nonlinear response saturates beyond a threshold power can be modeled by the standard NLSE. 
Under the slowly varying envelope approximation, the complex-valued electric field satisfies the following \textit{nonlinear Schr\"{o}dinger equation} (NLSE) \cite{agrawal2000nonlinear},
\begin{align}
	i\partial_\xi\psi+\sum_{m\geq 2}i^m\delta_m (\partial_\tau)^m\psi+(1+i\tau_{sh}\partial_\tau)f(|\psi|^2)\psi=0.
	\label{nlse}
\end{align}
Here, we perform the typical scaling $\psi \rightarrow A/\sqrt{P_0}$, $\tau\rightarrow(t-v_g/z)/t_0$, $\xi\rightarrow z/L_D$, where $L_D=t^2_0/|\beta_2|$ is the dispersion length,  $\delta_m=\beta_m/(m!|\beta_2|t^{m-2}_0)$, $\beta_m$ is the m$^{th}$ order dispersion coefficient at the carrier frequency $\omega_0$. 
The normalized self-steepening parameter, relevant for SDG fibers, is defined as, $\tau_{sh}=(\omega_0 t_0)^{-1}$ where $t_0$, $P_0$, $v_g$, $\gamma$ being the input pulse duration, input peak power, group velocity and nonlinear coefficient, respectively. For our proposed waveguide, the refractive index of the LN core at the operating wavelength ($\lambda_0 \approx 1.73$ $\mu$m) is around $n_\text{core}=2.13$ which offers a large refractive index contrast between core and silica cladding ($n_\text{clad}=1.44$), thereby resulting in tight mode confinement as indicated in the inset Fig.~\ref{fig1}. 
 For the input pulse width $t_{0} =55 $ fs, the normalized dispersion coefficients are calculated as $\delta_4=-0.0125$, and  $\delta_3= 0.0053$. Note, the value of $\delta_3$ is one order less than $\delta_4$.
For our system, we adopt the frequently used mathematical form of the SN response \cite{gatz1991soliton,hickmann1993modulational,hadvzievski2004power,raja2010modulational} as,
\begin{equation}
	f(|\psi|^2)=\frac{\mu|\psi|^2}{1+s|\psi|^2}, 
	\label{sn} 
\end{equation}
where  $\mu=\pm 1$ determines focusing($+$) or defocusing ($-$) nonlinearity.  The saturation parameter $s$ is defined as, $s=2|\beta_2|(t_0^2E_0k_0r_{ij}n_0^3)^{-1}$ which depends on the bias field ($E_0$), electro-optic coefficient ($r_{ij}$) and unperturbed  index of refraction ($n_0$) of the crystal in use \cite{christodoulides1995bright}. Note, being a PR crystal, LN offers additional tunability in controlling the saturable parameter $s$ through external bias field.  
For our proposed waveguide structure, in the proximity
of \textit{quartic point}, the governing equation for the field (see Eq. \eqref{nlse})
can be approximated as,
\begin{equation}
	i\partial_{\xi}\psi-\delta_2\partial^2_{\tau}\psi+\delta_4\partial^4_{\tau}\psi+\frac{\mu|\psi|^2}{1+s|\psi|^2}\psi =0
	\label{nnlse}
\end{equation}
Here the coefficients $\delta_2$ and $\delta_4$ are dominating and we neglect all higher order dispersion terms  ($\delta_{m >4}=0$) including the 3OD (as $|\delta_3/\delta_4|<<1$).  For PR crystals, we ignore the shock effect ($\tau_{sh}=0$) and the impact of two-photon absorption is minimal, as the energy of two-photon quanta ($\approx 0.7$ eV) is significantly lower than the energy band gap of LN ($\approx 3.9$ eV) \cite{beyer2005investigation}. Additionally, since experiments show no notable Raman shift \cite{lu2019octave,yu2019coherent} in LN, this effect is also excluded.The SN media can excite solitary waves in the form $\psi(\xi,\tau)=\sqrt{\psi_s(\tau)}e^{iq\xi}$ , where $q$ stands for propagation constant \cite{gatz1991soliton}. In our case, however, we seek for a stationary solution in the form $\psi(\xi,\tau)=g(\tau)e^{iq\xi}$ and substituting it in Eq.~\eqref{nnlse} we can have the equation for $g(\tau)$ as,
\begin{equation}
	-qg(\tau)-\delta_2\partial^2_{\tau}g(\tau)+\delta_4\partial^4_{\tau}g(\tau)+\frac{\mu g(\tau)^3}{1+sg(\tau)^2} =0. 
	\label{evg}
\end{equation}
As no internal energy flow in the solution, we can assume $ g(\tau)$ to be real. For a QS $ g(\tau)$ should have the specific form, $ g(\tau) = \mathcal{A}(q) \sech^2[\kappa(q)\tau]$ where $\mathcal{A}(q)$ and $\kappa(q)$ are related to the amplitude and width of the pulse. The amplitude-width relation which is the essential feature of the solution can be obtained by employing \textit{Ritz's optimization} procedure. The Lagrangian density ($\mathcal{L}$) corresponding to Eq.\eqref{evg} is expressed as,
\begin{equation}
	\mathcal{L}=\frac{1}{2}	\left[(\frac{\mu}{s}-q)g^2+\delta_2 |\partial_{\tau}g|^2+\delta_4 |\partial^2_{\tau}g|^2-\frac{\mu}{s^2}\ln(1+sg^2)\right]. 
	\label{eq:Lag}
\end{equation}
The static Lagrangian is obtained by employing the \textit{ansatz} function $g(\tau)=\mathcal{A}\sech^2(\kappa\tau)$ as, $L=\int_{-\infty}^{\infty} \mathcal{L} d\tau$ which follows,

\begin{equation}
	L =-\frac{2\mathcal{A}^2}{3\kappa} \left[q-\frac{4}{5}\delta_2\kappa^2- \frac{16}{7} \delta_4 \kappa^4\right]  + \frac{\mu}{s\kappa}\left[\frac{2}{3}\mathcal{A}^2-\frac{1}{2s}\Re[\Theta^2]\right],
	\label{eq:rLag}
\end{equation}
where, $\Re$ represents the real part, $\Theta=\cosh^{-1}\zeta$ with, $\zeta=1+2i\mathcal{A}\sqrt{s}$. Optimizing the static Lagrangian by  \textit{Euler-Lagrange} equation $\partial L/\partial j=0$ for  $j=\mathcal{A},\kappa$ we get an unique relationship between $\mathcal{A}$ and $\kappa$ (See details in Appendix \ref{appen_A}),

\begin{equation}
	\kappa=\left[\frac{-\Gamma_2+\sqrt{\Gamma_2^2-2\Gamma_4 \mathcal{F}}}{\Gamma_4}\right]^{1/2},
	\label{eq:ak}
\end{equation}
where, $\Gamma_2=32|\delta_2|/15$, $\Gamma_4=512|\delta_4|/21$ and   $\mathcal{F}=\frac{\mu}{\mathcal{A}^2s^2}\Re\left[\frac{\sqrt{\zeta-1}}{\sqrt{\zeta+1}}\Theta-\Theta^2\right]$, which is negative ($\mathcal{F}<0$) for the given range. For PQS ($\Gamma_2=0$) the relation reduces to $\kappa=(\sqrt{-2\mathcal{F}/\Gamma_4})^{1/2}$. 

\begin{figure}[h!]
 	\centering
 	\includegraphics[width=\linewidth]{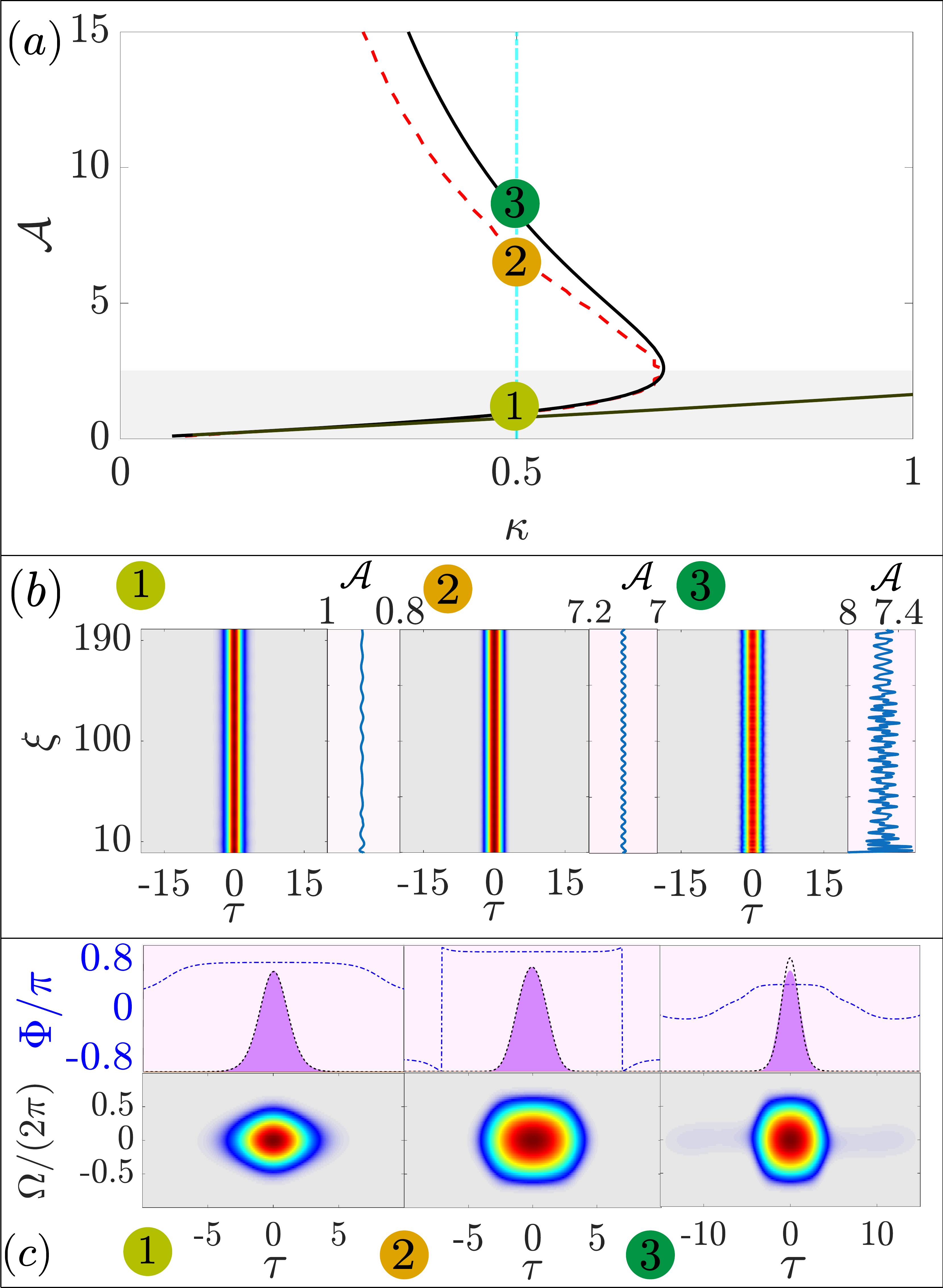}
 	\caption{($a$) Bistable $\mathcal{A}$-$\kappa$ relation derived analytically (solid black line) and by solving Eq. \eqref{evg} numerically (red dotted line) for $\delta_4=-0.0125$ and $s=0.5$. The gray line indicates the relation under the limit  $s \to 0$.
 		Plot $(b)$ represents the shape preserving dynamics of BQS corresponds to the points 1-3  on $\mathcal{A}$-$\kappa$ curve. In side panel the variation of peak amplitude is demonstrated.
 		In plot (c) we illustrate the output profile and temporal phase distribution across the pulse along with spectrograme in time ($\tau$) and frequency ($\Omega/2\pi$) space.  
 	}
 	\label{fig2}
 \end{figure}

In Fig. \ref{fig2} ($a$), the $\mathcal{A}$-$\kappa$ relation is illustrated for a saturation parameter of $s=0.5$ (solid line). This relation is derived analytically through the VA using a $\sech^2$ ansatz. Alternatively, the relation (red dotted line) is obtained through a numerical solution of Eq.~\eqref{evg} for $g(\tau)$, under the boundary condition $lim_{\tau \rightarrow \pm \infty} g(\tau) = 0$. The results from both methods are compared, showing good agreement, particularly in the lower branch indicated by a shaded area. However, a minor discrepancy appears in the upper branch due to deviations from the $\sech^2$ profile by the QS which tends towards a Gaussian shape. 
 
The discussed approaches reveal a bistable relationship between $\mathcal{A}$ and $\kappa$, indicating the existence of two possible QS states with the same pulse duration but varying amplitudes. This bistability, characteristic of a medium with SN \cite{gatz1991soliton,Gatz1992,Krolikowski1992} , has not been previously investigated in the context of QS. In the limit when $s$ approaches zero, as shown in Fig \eqref{fig2} ($a$) by gray line, the $\mathcal{A}$-$\kappa$ relation becomes non-bistable, indicating a reduction to pure Kerr-type nonlinearity, where bistable QS states do not exist. The mathematical formulation under this limit yields $lim_{s\rightarrow 0} \kappa(s)= \left[-\Gamma_2+\sqrt{\Gamma_2^2-2\Gamma_4 \chi \mathcal{A}^2}/\Gamma_4 \right]^{1/2}$, with $\chi=-16/35$, and aligns with previous studies \cite{karlsson1994soliton,roy2013formation}.

 In Fig.~\ref{fig2} ($b$), the shape-invariant propagation of QS is illustrated for three distinct points on the amplitude-width curve marked as \circled{1}-\circled{3}. Two bistable QS states are depicted with the same widths but varying peak amplitudes. A side panel presents the peak amplitude variation during propagation. Notably, in plot (b), the amplitude fluctuation is more significant at point  \circled{3} compared to \circled{2}, attributed to point \circled{2}'s alignment with a more precise $\mathcal{A}$-$\kappa$ relation obtained through numerical methods without any approximation.
 {Fig.~\ref{fig2} ($c$) displays the shape-preserving output profile alongside a uniform temporal phase distribution, reinforcing the solitonic behavior  and an XFROG-spectrogram diagram, emphasizing the QS's robustness. XFROG is a standard technique used to represent ultrashort pulses in time and frequency domain and is defined as the convolution $S(\tau,\omega,\xi)=|\int_\infty^\infty \psi(\xi,\tau')\psi_w(\tau-\tau')\exp(i\omega\tau') d\tau'|^2$, where $\psi_{w}$ is the reference window function.}  \\
 
\begin{figure}
     \centering
     \includegraphics[width=\linewidth]{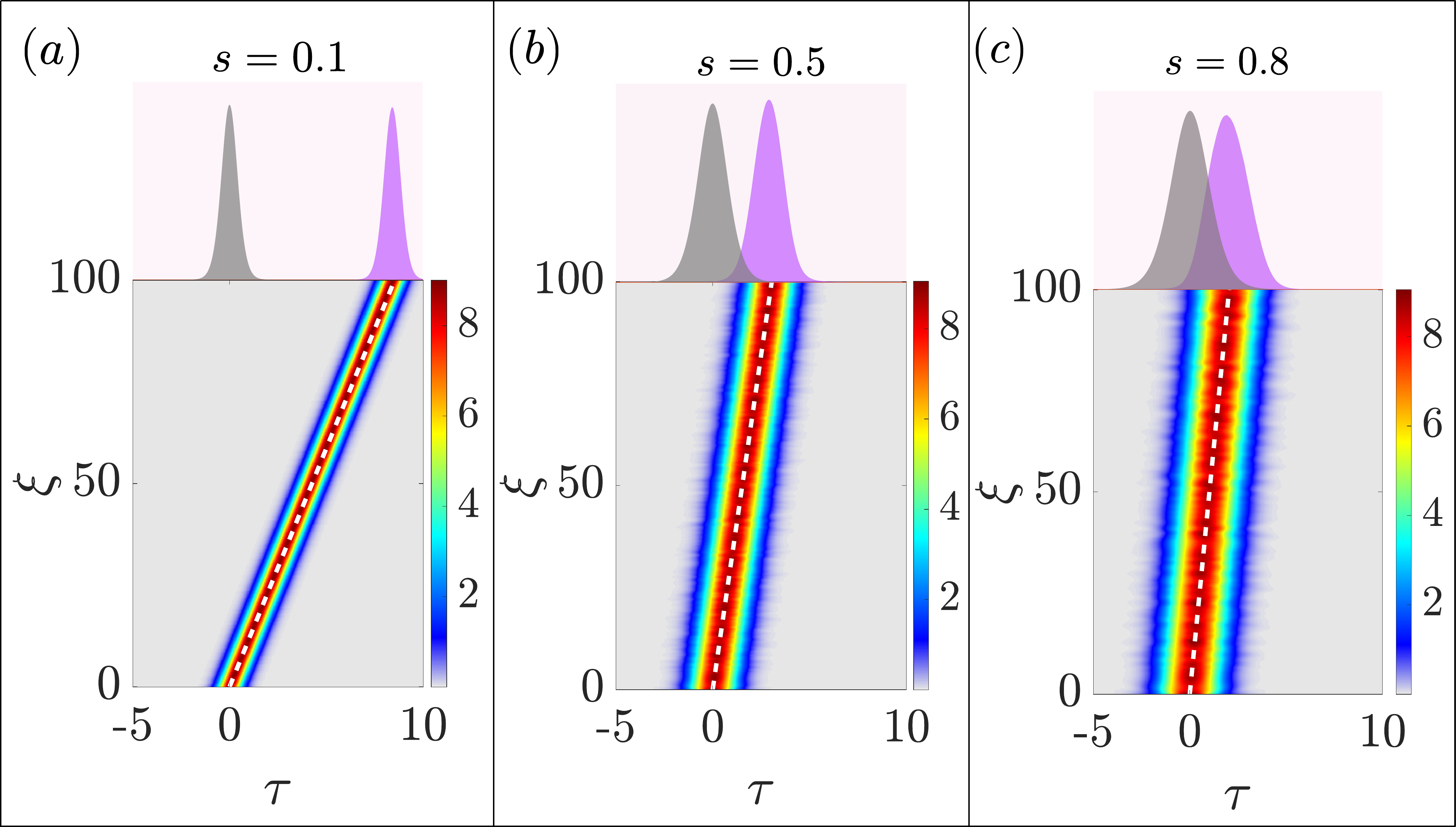}
     \caption{ Plot $(a)$-$(c)$ represent dynamics of QS under shock effect with the value $\tau_{sh}=0.017$ calculated for various $s$ values for $\delta_4=-0.0125$. The white dashed lines indicate the analytical prediction of the shock-mediated temporal shift of QS.}
     \label{fig:3}
 \end{figure}

 \textit{\textbf{Effect of self-steepening:}}   {The intensity-dependent index of refraction can distort an optical pulse during propagation, potentially leading to phenomena such as optical shocks due to pulse self-steepening. This effect arises where an increase in the index causes the trailing edge of the pulse to steepen, enabling the intensity to decline as rapidly as dispersion allows. This steepening resembles the development of an acoustic shock observed at the leading edge of a sound wave and occurs in materials where the peak of the pulse travels slower than its wings, allowing the trailing part to approach the peak.} 
 For the sake of completeness, we investigate the dynamics of QS under \textit{shock effect} which acts as a perturbation and influences the wave.  Fig.~\ref{fig:3} ($a$)-($c$) highlight the robust propagation of QS across two branches under shock conditions, with a noted temporal position shift that characterizes the shock phenomenon \cite{agrawal2000nonlinear}. A variational analysis is employed to express this temporal position shift, denoted as $\Delta \tau_w$, considering the shock as a perturbation (See details in Appendix \ref{appen_B}).    
\begin{equation}
	\Delta \tau_w=\mathcal{A}_0^2\tau_{sh} \digamma_s \xi,
	\label{eq:shock}
\end{equation}
where a hyperbolic secant function is considered as an ansatz and the parameter $\digamma_s$ is defined as, $\digamma_s=\frac{1}{m}-\frac{2}{m}\frac{\ln(\sqrt{m}+\sqrt{m+1})}{\sqrt{m}\sqrt{m+1}}-\left[\frac{\cos^{-1}(1+2m)}{2m}\right]^2$ with $m=s\mathcal{A}_0^2$ (see details in Appendix \ref{appen_B}), where $\mathcal{A}_0$ represents the input amplitude. In Fig.~\ref{fig:3} ($a$)-($c$), the white dotted lines indicate the analytical prediction (based on Eq. \eqref{eq:shock}) of temporal-shift which shows an excellent agreement with the full numerical result.\\ 

\begin{figure}
    \centering
    \includegraphics[width=\linewidth ]{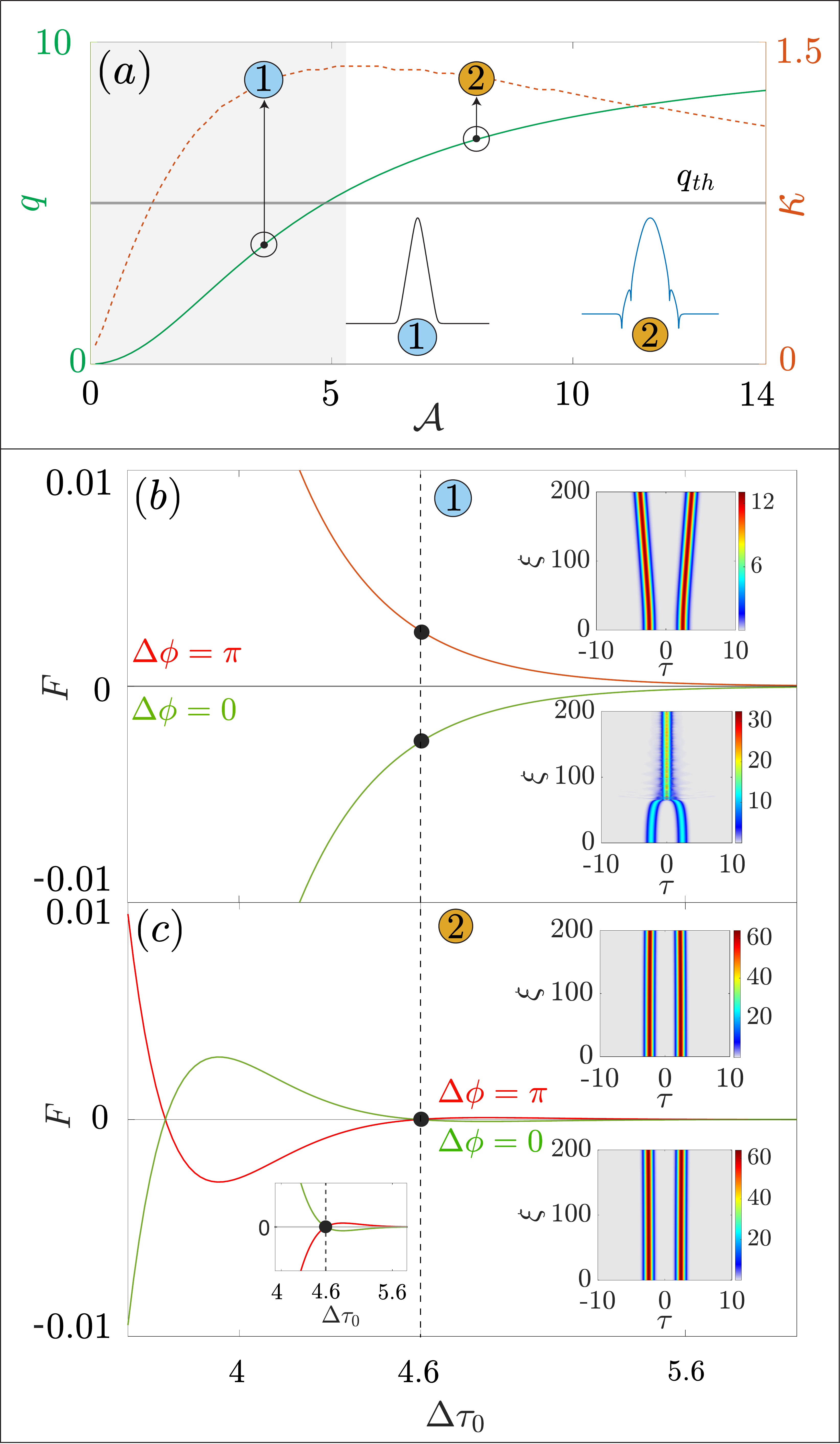}
    \caption{(a) Propagation constant $q$ as a function of $\mathcal{A}$ (solid line) and $\mathcal{A}$-$\kappa$ relation for $\delta_2=-1/2$, $\delta_4=-0.0125$ and $s=0.1$. The horizontal solid line indicates $q_{th}$. The inset displays the soliton shape at the lower and upper branches, indicated by 1 and 2, respectively.
    (b) Force curve as a function of separation $\Delta \tau_0$ for in phase ($\Delta \phi=0$) and out-of-phase ($\Delta \phi=\pi$) BQS pair in lower branch ($q<q_{th}$). In the inset we show repulsive (for $F>0$) and attractive (for $F<0$) dynamics of BQS pair for identical separation $\Delta \tau_0=4.6$.
    (c) Force curve as a function of  $\Delta \tau_0$ for in phase ($\Delta \phi=0$) and out-of-phase ($\Delta \phi=\pi$) BQS pair in higher branch ($q>q_{th}$). The force curves exhibit a fixed point ($F=0$) at the same separation $\Delta \tau_0=4.6$, and leads to the formation of a bound-state.}
    \label{fig:44}
\end{figure}

\textit{\textbf{Interaction of BQS:}}
{This subsection examines the BQS interaction. In contrast to PQS, the general QS displays an oscillatory tail when the propagation constant is more than the threshold value $q_{th}=\delta_2^2/4|\delta_4|$ \cite{akhmediev1994radiationless}. Through Ritz's optimization of the static Lagrangian, the propagation constant $q$ is expressible as a function of amplitude $\mathcal{A}$ (See details in Appendix \ref{appen_A}).
\begin{equation}
	q=\frac{3\kappa^2}{8} \left(\Gamma_2+\frac{3 \Gamma_4\kappa^2}{4} \right)+\frac{\mu}{s}\left(1-\frac{3}{4\mathcal{A}^2s} \Re{[\Theta^2]} \right).
	\label{eq:q}
\end{equation} 
The propagation constant $q$ of a QS may change with amplitude and can cross the threshold value \( q_{th} \), leading to the emergence of an oscillatory tail. When two QSs exhibit such exponentially decaying tails, they can interact to create a bound state \cite{Deng2025} owing to tail overlap. The relationship, including the propagation constant \( q \) versus amplitude \( \mathcal{A} \) and the corresponding \( \mathcal{A} \)-\( \kappa \) dynamics, is represented in \figref{fig:44} ($a$) for a saturation parameter \( s=0.1 \). Two scenarios are considered: one, for the value of propagation constant below \( q_{th} \) indicated by  \circled{1}, and another above it indicated by  \circled{2}. The BQS generated for $q>q_{th}$ may form bound state based on the weak interaction forces induced by the tail-tail overlapping. Exploiting the concept of momentum $P=\int_{-\infty}^{\infty} \frac{i}{2}\left(\psi\psi^*_{\tau}-\psi^*\psi_{\tau}\right) d\tau$ it is possible to calculate the interaction force as, $F=\frac{dP}{d\xi}$. For exponentially decaying tail ($q<q_{th}$) the force $F$ can be calculated as (see Appendix \ref{appen_c} for details), 
\begin{equation}
    F=-g_0^2n^2e^{i\Delta\phi}(2q-n^2\delta_2+3n^4\delta_4)e^{-2n\tau_0}.
    \label{F1}
\end{equation}
For a decaying oscillatory tail, we can form $F$ as,
{\small
\begin{equation}
\begin{split}
F=g_0^2 e^{i\Delta\phi}e^{-2 G\tau_0}\left[ \delta_2(G \cos\alpha+ K \sin\alpha)^2 
-\frac{\delta_4}{2}[(4G^4-H) \right.
\\
\left.-3(4G^2K^2-H)\cos2\alpha +12GK\sqrt{H}\sin2\alpha]-2q\cos^2\alpha \right]   
    \end{split}
    \label{F2}
\end{equation}
}
where, $H=(G^2-K^2)^2$ and $\alpha=(K\tau_0+\theta)$. The values of $g_0$ and $\theta$ can be determined by fitting the soliton tail with $g_0e^{-G\tau}\cos\alpha$.} 
{The force \( F \) between soliton pairs can either attract or repel depending on it's sign, with zero force suggesting possible bound state formation. As illustrated in \figref{fig:44} ($b$), for \( q < q_{th} \), in-phase solitons attract while out-of-phase solitons repel, but no bound state forms as the force decreases to zero asymptotically. Conversely, for \( q > q_{th} \), as shown in \figref{fig:44} ($c$), the force curve exhibits a fixed point where \( F = 0 \), indicating a bound state. As predicted by the force analysis, particularly at a separation \( \Delta \tau_0 = 4.6\),  a bound state appears for both in-phase and out-of-phase QS pairs.} 

\section{Stability Analysis}
Finally, we investigate the stability of BQS in saturable media through \textit{linear stability analysis} (LSA). In this approach, the stationary QS, denoted as \(g(\tau)\), is subjected to small amplitude perturbations represented by \(v(\tau)\) and \(w(\tau)\). The optical field is expressed as \(\psi(\tau,\xi)=\left[g(\tau) +w^{*}(\tau)e^{h^{*}\xi}+v(\tau)e^{h\xi}\right]e^{iq\xi}\). By substituting this form into the governing equation  Eq. \eqref{nnlse} and performing linearization with respect to the perturbations, we derive two linear systems for \(v\) and \(w\) that yield the growth rate \(h\). This process culminates in an eigenvalue problem as follows,
\begin{equation}
	\mathcal{O}X=hX
\end{equation}
where, $X=[v,w]$ and the matrix operator $\mathcal{O}$ is,
	\begin{align}
	\mathcal{O} = i\begin{bmatrix}
		\tilde{\nabla}+\alpha_0 & \alpha_1 \\
		-\alpha_1 & -(\tilde{\nabla}+\alpha_0)
	\end{bmatrix}.
\end{align}
The parameters are defined as, $\tilde{\nabla}=\delta_4\partial^4_\tau-\delta_2\partial^2_\tau$, $\alpha_0=-q+\mu g^2\mathcal{G}(1+\mathcal{G})$, $\alpha_1=\mu g^2\mathcal{G}^2$, with $\mathcal{G}=(1+sg^2)^{-1}$.  
We utilize the Fourier collocation method \cite{yang2010nonlinear} to address the eigenvalue problem and obtain the entire spectrum of the linear-stability operator $\mathcal{O}$. An eigenvalue with a positive real part, $\Re(h)>0$, signifies that the perturbed stationary solution is unstable. We compute the LSA spectra by examining how $\Re(h)>0$ varies with pulse amplitude ($\mathcal{A}$) and saturation parameter ($s$), while maintaining a constant 4OD coefficient.

\begin{figure}[h!]
	 	\centering
	 	\includegraphics[width=\linewidth]{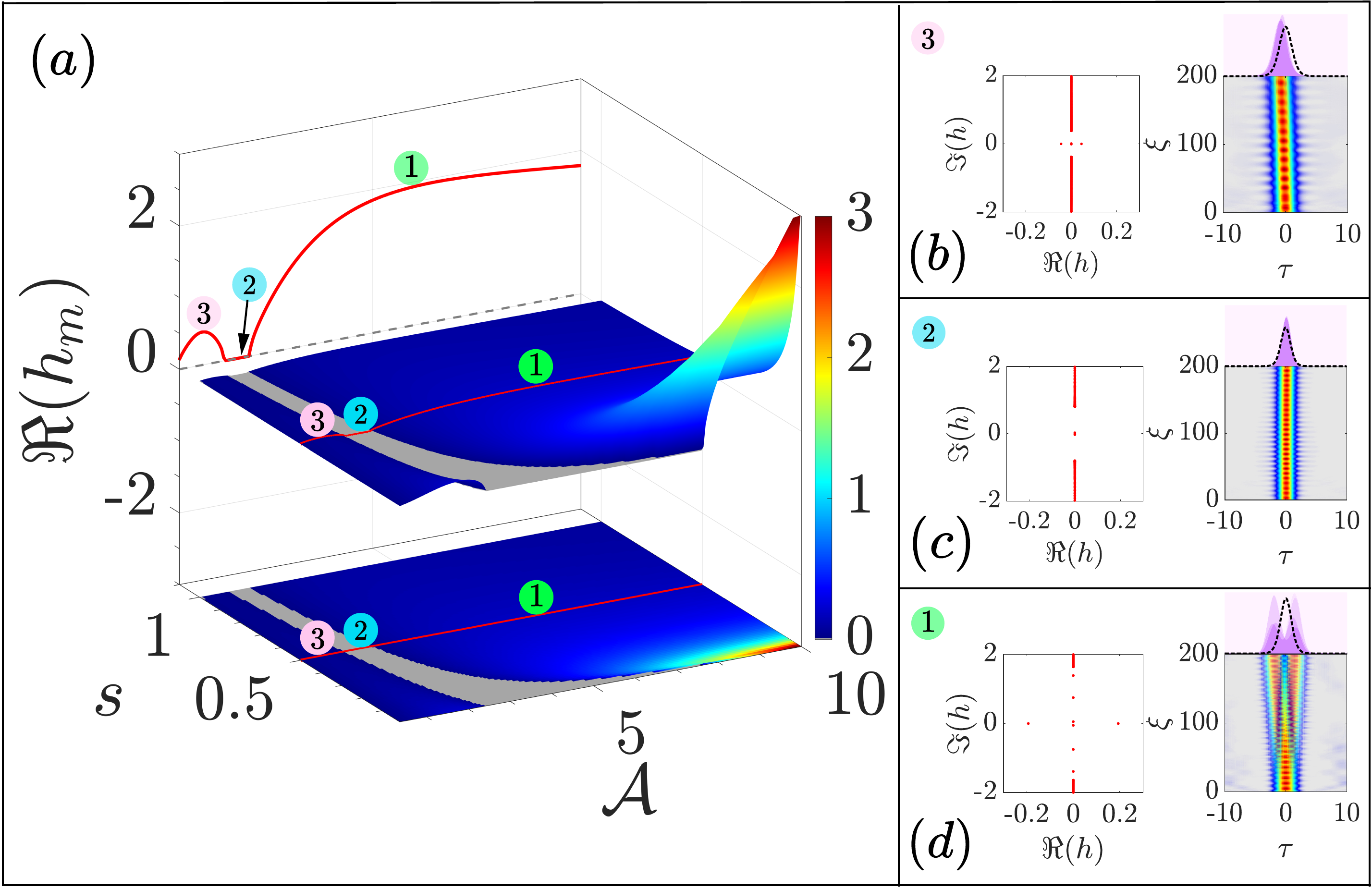}
	 	\caption{$(a)$ Instability phase diagram in $\mathcal{A}$-$s$ space. Plot $(b)$-$(d)$ demonstrate the propagation of QS corrupted by amplitude noise along with eigen-value spectra of instability growth $\Re(h_{m})$ for $s=0.5$ and $\delta_4=-0.0125$. The numeric values of the pulse parameters are for point 1, $\mathcal{A}=6$, $\kappa=0.58$  $q=1.65$ (see plot $(d)$), for point 2,  $\mathcal{A}=1.78$, $\kappa=0.66$, $q=0.8$ (see plot $(c)$) and for
	 	point 3,  $\mathcal{A}=1$, $\kappa=0.52$ $q=0.38$ (see plot $(b)$).  }
	 	\label{fig3}
	 \end{figure}

   	The stability phase plot (see Fig. \ref{fig3}) in the $\mathcal{A}$-$s$ parametric space reveals that BQSs are stable within a limited region where the maximum instability eigenvalue, $\Re[h_{max}]$, is zero. To confirm the LSA findings, QS stability was examined by introducing a $10\%$ noise in pulse amplitude at three distinct settings. Utilizing the \textit{Crank-Nicolson algorithm}, the evolution of the pulse was analyzed for $s=0.5$ and $\delta_4=-0.0125$. The results indicated that the QSs were linearly unstable at points \circled{1} ($\mathcal{A}=6$, $s=0.5$) and \circled{3} ($\mathcal{A}=1$, $s=0.5$), whereas they exhibited stability at point \circled{2} ($\mathcal{A}=1.78$, $s=0.5$). The linear-stability spectra revealed that points \circled{1} and \circled{3} contained positive real part eigenvalues, confirming linear instability, in contrast to point \circled{2}, where all eigenvalues were imaginary, indicating linear stability. Additionally, in all cases, the continuum eigenvalue edges  at $\Im(h)=\pm q$ and the pair of discrete eigenvalues on the imaginary axis indicates the internal modes contributing to shape oscillations in the soliton pedestal \cite{yang2010nonlinear}.
   
	In summary, we demonstrate the existence of novel BQS within an optical medium characterized by saturable nonlinearity. Through a theoretical analysis employing variational optimization, we find that unique bistable solitonic states with varying amplitude profiles yet identical widths can persist. A realistic waveguide is engineered using the LN crystal, promoting an advantageous dispersion environment for BQS excitation. A perturbative variational analysis is also employed to grasp the influence of the optical shock on BQS that arises due to the intensity-dependent group velocity. {Further, we investigate the interaction of the BQS pair for various conditions and by exploiting the force analysis demonstrate the formation of bound state.}
    Finally, a linear stability analysis is conducted to confirm the stability of this unique structure against perturbations. Our findings enhance the understanding of self-organized temporal structures as they develop under specific dispersion environment in saturable nonlinear media, indicating potential applications in high-power laser systems and communication technologies.

	\section*{Acknowledgment}   
	T.D. is financially supported by IIT Kharagpur, the Ministry of Education of the Government of India, 
	and A.P.L. is supported by Anusandhan National Research Foundation (ANRF) erstwhile SERB, Government of India.

\appendix
\section{Derivation of the parametric relationship}\label{appen_A}
The static Lagrangian in reduced form is given in Eq. \ref{eq:Lag}.  By adopting Euler Lagrangian equation $\frac{\partial}{\partial \xi}\frac{\partial L}{\partial p_{\xi}}-\frac{\partial L}{\partial p}=0$ for $p=\mathcal{A},\kappa$ we get two equations, 
\begin{equation}
\begin{split}
    \frac{\partial L}{\partial \mathcal{A}}= -\frac{4\mathcal{A}q}{3\kappa}+\frac{16\kappa\mathcal{A}\delta_2}{15}+\frac{64\kappa^3\mathcal{A}\delta_4}{21}+\frac{4\mu\mathcal{A}}{3s\kappa}\\-\frac{2\mathcal{A}\sqrt{s}\mu}{\kappa s^2}\Re\left[\frac{i\Theta}{\sqrt{\zeta^2-1}}\right]=0
    \end{split}
\label{Lag_A}    
\end{equation}
\begin{equation}
\begin{split}
    \frac{\partial L}{\partial \kappa}= \frac{2\mathcal{A}^2q}{3\kappa^2}+\frac{8\mathcal{A}^2\delta_2}{15}+\frac{96\kappa^2\mathcal{A}^2\delta_4}{21}-\frac{2\mu\mathcal{A}^2}{3s\kappa^2}\\+\frac{\sqrt{s}\mu}{2\kappa^2 s^2}\Re\left[\Theta^2\right]=0
    \end{split}
\label{Lag_k}    
\end{equation}
where $\zeta=1+2i\sqrt{s}\mathcal{A}$ and, $\Theta=\cosh^{-1}(\zeta)$.
Eliminating the parameter $q$ from Eq.\ref{Lag_A} and \ref{Lag_k} we have,
\begin{equation}
   \kappa^4+2\Gamma_2\kappa^2/\Gamma_4+\mathcal{F}/\Gamma_4=0
    \label{quad_Ak}
\end{equation}
where,  $\Gamma_2=32|\delta_2|/15$, $\Gamma_4=512|\delta_4|/21$ and   $\mathcal{F}=\frac{\mu}{\mathcal{A}^2s^2}\Re\left[\frac{\sqrt{\zeta-1}}{\sqrt{\zeta+1}}\Theta-\Theta^2\right]$.
Solution of  Eq.\ref{quad_Ak} leads to Eq.\ref{eq:ak}. The expression of $q$ can be derived from Eq.\ref{Lag_k} as expressed in Eq.\ref{eq:q}.\\

\section{Variational analysis under shock effect}\label{appen_B}
Under the perturbation of shock effect, the governing equation for QS in saturable nonlinearity is given as,
\begin{equation}
    i\partial_{\xi}\psi-\delta_2\partial^2_{\tau}\psi+\delta_4\partial^4_{\tau}\psi+\frac{\mu|\psi|^2}{1+s|\psi|^2}\psi=i\epsilon(\psi),
    \label{nlsep}
\end{equation}
where, the perturbation due to shock is taken account through $\epsilon(\psi)$  as,  $\epsilon(\psi)=\tau_{sh} \mu\frac{\partial}{\partial \tau}\frac{|\psi|^2\psi}{1+s|\psi|^2}$. The Lagrangian density of Eq. \ref{nlsep} is presented as, 
\begin{equation}
\begin{split}
    \mathcal{L}=\frac{i}{2}(\psi^*\psi_{\xi}-\psi^*_\xi\psi)+\delta_2|\partial_\tau\psi|^2+\delta_4|\delta^2_\tau\psi|^2+\\\frac{\mu}{s}\left[|\psi|^2-\frac{1}{s}\ln(1+s|\psi|^2)\right]+i(\epsilon^*\psi-\psi^*\epsilon).
    \end{split}
    \label{Lagp}
\end{equation}
Next Lagrangian density is reduced as $L=\int_{-\infty}^{\infty} \mathcal{L} d\tau$, by using the \textit{ansatz}  function $\psi=\mathcal{A}\sech[\kappa(\tau-\tau_w)]e^{i[\phi-\delta(\tau-\tau_w)]}$. Note, all the parameters $\mathcal{A}$, $\kappa$,$\phi$,$\delta$ and $\tau_w$ are now function of $\xi$. The reduced Lagrangian takes the form,
\begin{equation}
\begin{split}
   L= \frac{2\mathcal{A}^2}{\kappa}\left(\phi_\xi+\delta\tau_{w\xi}\right)+\delta_2\left(\frac{2\mathcal{A}^2\delta^2}{\kappa}+\frac{4\mathcal{A}^2\kappa}{15}\right)\\+\frac{2\mathcal{A}^2\delta_4}{\kappa}\left(\delta^4+2\delta^2\kappa^2+\frac{7\kappa^4}{15}\right)\\+\frac{\mu}{s}\left(\frac{2\mathcal{A}^2}{\kappa}-\frac{1}{2s\kappa}\left[\cosh^{-1}(1+2s\mathcal{A}^2)\right]\right)\\+\int_{-\infty}^{\infty} (\epsilon^*\psi-\psi^*\epsilon) d\tau,
   \end{split}
    \label{rLagp}
\end{equation}
where, $\phi_\xi$ and $\tau_{w \xi}$ represent the derivative with respect to $\xi$. Next by employing Euler Lagrangian equation $\frac{\partial}{\partial \xi}\frac{\partial L}{\partial p_{\xi}}-\frac{\partial L}{\partial p}=0$ for $p=\delta,\tau_w,\phi$, we get,
$\frac{\partial}{\partial \xi}\left(\frac{2\mathcal{A}^2}{\kappa}\right)=0$ , $\frac{\partial \delta}{\partial \xi}=0$ and the dynamic equation for temporal position emerges as, 
\begin{equation}
    \frac{\partial \tau_w}{\partial \xi}=-2\delta\delta_2-4\delta^3\delta_4-2\delta_4\delta\kappa^2+\tau_{sh}\mathcal{A}^2\digamma_s,
    \label{tawx}
\end{equation}
where, the parameter $\digamma_s=\frac{1}{m}-\frac{2}{m}\frac{\ln(\sqrt{m}+\sqrt{m+1})}{\sqrt{m}\sqrt{m+1}}-\left[\frac{\cos^{-1}(1+2m)}{2m}\right]^2$ with $m=s\mathcal{A}^2$ . Now setting $\delta=0$ (no initial frequency) in Eq. \ref{tawx},  we can obtain Eq. \ref{eq:shock}.
\section{Analytical formalism of two soliton interaction force}\label{appen_c}

Inspired by the pioneering work by Manton \cite{Manton1979}, the soliton interaction can be addressed by defining the total momentum of the field as \cite{Deng2025},  
\begin{equation}
    P=\int_{-\infty}^{\infty} \frac{i}{2}\left(\psi\psi^*_{\tau}-\psi^*\psi_{\tau}\right) d\tau. 
\end{equation}
The interaction force can be calculated as, $F=\frac{dP}{d\xi}$. For a stationary field  $\psi=g(\tau)e^{iq\xi}$ in the interval $\left[\tau_1,\tau_2\right]$, the force becomes, 
 \begin{equation}
 \begin{split}
     F=\left[\frac{\delta_2}{2}g_{\tau}^2+\frac{1}{2}\delta_4\left(g_{\tau\tau}^2-g_{\tau}g^*_{\tau\tau\tau}-g_{\tau\tau\tau}g^*_{\tau}\right) \right.
     \\
     \left. -\frac{\mu}{2s}(g^2-\frac{1}{s}\ln(1+sg^2))+qg^2\right]^{\tau_2}_{\tau_1}
     \end{split}
     \label{C2}
 \end{equation}
 The double-hump soliton $g(\tau)=\eta(\tau+\tau_0)+\eta(\tau-\tau_0)e^{i\Delta\phi}$  consists of two BQSs $\eta(\tau\pm\tau_0)$. The force exerted on the trailing soliton by the leading soliton can be realized by setting $\tau_1=0$ and $\tau_2=\infty$, and by adopting the condition $g(\tau)\rightarrow 0$ for $\tau_2\rightarrow \infty$, we get,
\begin{equation}
 F\approx-\left[\frac{\delta_2}{2}g_{\tau}^2+\frac{1}{2}\delta_4\left(g_{\tau\tau}^2-g_{\tau}g^*_{\tau\tau\tau}-g_{\tau\tau\tau}g^*_{\tau}\right)+qg^2\right]_{\tau_1=0}.
 \label{C3}
 \end{equation}
Now neglecting the effect of self-interacting terms from Eq. \ref{C3} we may obtain,
 \begin{equation}
 \begin{split}
     F\approx -e^{i\Delta\phi}\left[\delta_2\eta_{1\tau}\eta_{2\tau}+\delta_4\left(\eta_{1\tau\tau}\eta_{2\tau\tau}-\eta_{1\tau}\eta_{2\tau\tau\tau}\right.\right.
     \\ \left.\left. -\eta_{2\tau}\eta_{1\tau\tau\tau}\right) 
      +2q\eta_1\eta_2 \right]_{\tau_1=0}
      \end{split}
     \label{C4}   
 \end{equation}
 where the trailing tail of the leading soliton and the leading tail of the trailing soliton are defined as, $\eta_1$ and $\eta_2$, respectively. For BQS tail, Eq. \ref{evg} can be modified as,
\begin{equation}
    -qg(\tau)-\delta_2\partial^2_{\tau}g(\tau)+\delta_4\partial^4_{\tau}g(\tau)=0
    \label{C5}
\end{equation}
now setting tailing function with the form $g(\tau)=g_0e^{n\tau}$  from Eq. \ref{C5} we estimate the $n$ as,
\begin{equation}
    n=\pm\sqrt{\frac{|\delta_2|\pm\sqrt{|\delta_2|^2-4q|\delta_4|}}{2|\delta_4|}}
    \label{C6}
\end{equation}
Now for the decaying tail where $\Im(n)=0$, $\eta_{1,2}$ takes the form, $\eta_1=g_0e^{-n(\tau+\tau_0)}$ and $\eta_2=g_0e^{n(\tau-\tau_0)}$.  Exploiting Eq. \ref{C4}, the expression of force $F$ for decaying tail can be obtain as,
\begin{equation}
    F=-g_0^2n^2e^{i\Delta\phi}(2q-\delta_2n^2+3n^4\delta_4)e^{-2n\tau_0}
    \label{C7}
\end{equation}
Now for $q>q_{th}$, $\Im(n)\ne 0$, and we have a decaying oscillating tail.  For such condition according to \cite{Deng2025}, $\eta_{1,2}$ takes the form $\eta_1=g_0e^{-G(\tau+\tau_0)}\cos[K(\tau+\tau_0)+\theta]$ and $\eta_2=g_0e^{G(\tau-\tau_0)}\cos[K(\tau-\tau_0)-\theta]$ where $G$ and $K$ are the real and imaginary part of $n$. 
Utilizing Eq. \ref{C4} the expression of force can be calculated as,
{\small
\begin{equation}
\begin{split}
F=g_0^2 e^{i\Delta\phi}e^{-2 G\tau_0}\left[ \delta_2(G \cos\alpha+ K \sin\alpha)^2 
-\frac{\delta_4}{2}[(4G^4-H) \right.
\\
\left.-3(4G^2K^2-H)\cos2\alpha +12GK\sqrt{H}\sin2\alpha]-2q\cos^2\alpha \right]   
    \end{split}
    \label{C8}
\end{equation}
}
here, $H=(G^2-K^2)^2$ and $\alpha=(K\tau_0+\theta)$. The values of $g_0$ and $\theta$ can be determined by fitting the soliton tail with $g_0e^{-G\tau}\cos\alpha$.  For \figref{fig:44}($b$)  the value of $n$ can be derived from Eq.(\ref{C6}) as, $n=3.13$, with $q=3.701$. The parameter $g_0$ is derived by fitting the tail with the function $g_0 e^{-n|\tau|}$ and we obtain $g_0\approx25$. For \figref{fig:44}($c$) the fitting function is $g_0 e^{-G\tau}\cos\alpha$ and we obtain the value of the parameters are, $g_0\approx69$, $G\approx4.11$, $K\approx2.578$ and  $\theta\approx-10.2$ with $q=6.96$. 
\bibliography{myref}
\end{document}